\documentclass[aps,two column,prl]{revtex4}

\begin{document}

\title{Comment on ``How the result of a single coin toss can turn out to be 100 heads"}

\author{Yakir Aharonov}
\affiliation{School of Physics and Astronomy, Tel Aviv University, Tel Aviv 6997801, Israel, \\ and Schmid College of Science, Chapman University, Orange, CA 92866}

\author{Daniel Rohrlich}
\affiliation{Department of Physics, Ben Gurion University of the Negev, Beersheba
8410501 Israel}

\date{\today}

\maketitle

The ``classical weak value" of Ferrie and Combes in a recent Letter \cite{fc} is not an analogue of a weak value.  It might better be characterized as a parody of a weak value.

	To understand the error in Ref. \cite{fc}, consider Bohm's formulation of quantum mechanics \cite{bohm} applied to a particle in a double-slit interferometer. The particle responds classically to a fluctuating ``pilot wave".  Is this formulation a classical analogue of quantum interference?  No, because the pilot wave is quantum; in particular, it has a wave number $k=p/\hbar$, where $p$ is the particle's momentum.  Quantum mechanics constrains the pilot wave.  Likewise, quantum mechanics constrains the errors in a weak measurement; they are not analogous to the noise in Ref. \cite{fc}.

	Weak values describe quantum systems subject to complete initial and final boundary conditions.  The resulting pre- and postselected ensemble (PPS) has no classical analogue, because complete initial and final boundary conditions on a classical system would be either redundant or inconsistent.  For a quantum system, inconsistency could arise only with orthogonal pre- and postselected states.  The pre- and postselected states chosen by Ferrie and Combes are indeed orthogonal (via noise) and thus irrelevant to weak values.

	How can we verify that initial and final boundary conditions characterize a PPS ensemble over the entire intermediate interval?  A measurement of an intermediate observable $A$ generally disturbs the ensemble; the disturbance is inherent and has no classical analogue.  Hence, although Ferrie and Combes state that sources of classical noise ``can be provided", they do not provide any classical analogue.  With refreshing candor, they attribute their measurement noise to an observer who has no time to recheck results, wears smudged glasses, or lies.  By contrast, the scatter in a weak measurement is the price we pay to limit the disturbance of the PPS ensemble.  We model the measuring device via a variable $Q_d$, the position of a pointer on a dial, and an interaction Hamiltonian $H_{int} = g(t) A P_d$, where $g(t)$ is a coupling and $P_d$ is conjugate to $Q_d$.  To limit the disturbance arising from the measurement, we must limit the range of $P_d$, hence bound $\Delta P_d$.  But since $\Delta Q_d \ge \hbar /2\Delta P_d$, this bound implies scatter in $Q_d$, i.e. in the measurement of $A$.

	We have seen three essential elements of weak values.  First, measurements on a PPS ensemble yield a more complete description of the intermediate reality than measurements on a preselected ensemble.  Second, the inherent scatter in measurements arises, via the uncertainty principle, from the measurement coupling, which must be weak so as not to disturb the PPS ensemble.  Neither element has a classical analogue.  The third essential element is the measuring device.  Although the final state of the {\it ensemble} is postselected, the final state of the {\it measuring device} is not.  Yet the values measured on the PPS ensemble cluster around the weak value, with the spread characteristic of measurements made with this device, as they would for any normal result.  The measuring device speaks for itself, just as clocks and rulers speak for themselves about relativistic kinematics, and screens, slits and springs speak for themselves about quantum limits on measurement \cite{bohr}.  

This third element of weak values, too, is absent from Ref. \cite{fc}.  If there is a measuring device in Ref. \cite{fc}, it is Bob, who reports the value of a variable $s$ taking values $\pm 1$.  The ``classical weak value", Eq. (35) in Ref. \cite{fc}, is not recorded by any measuring device.  In particular, consider the example in which ``the result of a single coin toss" turns out to be ``100 heads":  $\delta$ = 0.99, so the probabilities of $s = \pm 1$ are nearly equal, and the probability that Bob flips the coin or lies about the outcome is at most 0.01.  Since Alice postselects $s = -1$, Alice and Bob obtain at most one head for every coin toss.  Nothing even slightly anomalous here---except Eq. (35).

	To conclude, the ``anomalous" values of Ferrie and Combes say nothing about quantum---or even classical---physics.  They are not analogues of the weak values that emerge when we describe the quantum world via an initial state evolving forwards in time and a final state evolving backwards in time, and couple this world weakly to realistic measuring devices.
\\

Y.A. thanks the Israel Science Foundation, the ICORE Excellence Center ``Circle of Light", and the German-Israeli Project Cooperation (DIP) for support.  D.R. thanks the John Templeton Foundation and the Israel Science Foundation for support.

\end{document}